# Production of Entangled X-Ray Photon Pairs by High Energy Electrons

K.A. Ispirian, M.K. Ispiryan

Yerevan Physics Institute, Brothers Alikhanian 2, Yerevan, 0036, Armenia

**Abstract**

Without developing a correct theory, the rate of the production of entangled photon pairs in the X-ray region by high energy electrons is estimated for experimental parameters used in a few experiments on X-ray parametric down conversion. Taking into account the need for X-ray entangled X-ray photons, it is proposed to begin the experimental study of such mechanism of nonlinear X-ray optics before the launch of the X-ray free electron lasers.

The parametric down conversion (PDC) of a "pump" photon with frequency $\omega_p$ into two photons, a "signal" photon with $\omega_s$ and an "idler" photon with $\omega_i$, in a non-linear medium, provides two entangled photons. In the optical region, PDC, which is one of the most important processes of the nonlinear X-ray optics, has been theoretically and experimentally studied in many works. PDC can find wide scientific applications in quantum computers, cryptography, teleportation, etc. In X-ray region PDC for the first time has been considered in [1] and observed in [2]. Since the cross section of PDC is small, it has been further experimentally studied only in a few works [3-6] without satisfactory development and comparison between the theory and experiment. The highest event rate observed for the third generation SR sources [4-6] is ~ 0.1 Hz, while the expected one for LCLS and TESLA FELs is 10000 times higher [7]. As the optical, the X-ray PDC semiclassically is interpreted as mixing of the pump photon with a zero point fluctuation photon (ZPF), and the very small X-ray nonlinear susceptibilities are somewhat compensated by higher ZPF fields. The X-ray PDC takes place as non-linear Bragg diffraction in crystals with energy and momentum conservation laws

$$\omega_p = \omega_s + \omega_i, \quad \vec{k}_p + \vec{H} = \vec{k}_s + \vec{k}_i,$$

where $\vec{k}_j$ ($j=p,s,i$) are the corresponding momenta of the photons and $\vec{H}(j,k,l)$ is the crystal reciprocal lattice vector for the indices $j,k,l$.

The demand for intense source of entangled X-ray photon pairs for applications in various fields is so high that even such exotic processes as Unruh radiation [8] and inverse double Compton scattering [9] are proposed for production of entangled X-ray photon pairs. Postponing the development of the correct theory of photon entangled pair production by electrons (PEPPE) and using the results obtained for X-ray PDC [1,10], in this short paper we shall make simple estimates of the corresponding PEPPE yields for arrangement parameters coinciding with those of the works [1,4] in order to accelerate the development of the correct theory and experiments.

According to X-ray PDC theory [1] in the approximation $|\vec{k}_S| \approx |\vec{k}_i|$, when the phase matching is optimal, the number $N_S$ of the PDC signal photons produced by a pump photon and emitted into a solid angle $\Delta \Omega_S$ is equal to

$$\Delta N_S(\omega_P, \omega_S) = A(\omega_P, \omega_S) \Delta \Omega_S, \tag{1}$$

where

$$A(\omega_P, \omega_S) = \frac{8\pi \omega_S^3 (\omega_P - \omega_S)(\hbar \omega_P)}{c^4 [1 - \cos \theta_{Si}]} \left| \frac{G(hkl)}{V} \right|^2 \vec{\theta}_{spi}^2 (hkl) L_{eff}, \tag{2}$$

$$G(hkl) = \sum_n g_n(hkl) \exp[-2\pi i (hu_n + kv_n + lw_n)] \tag{3}$$

is the nonlinear structure factor, $g_n(hkl)$ are coefficients of the Fourier expansion of the linear susceptibility $\chi(\omega_P)$, $V$ is the volume of unit cell, $\vec{\theta}(spi)$ is a vector determining the polarization and depends on the experiment's geometry (see Eq. (11) of [1]), $L_{eff}$ is the crystal effective length, which for Bragg geometry is equal to

$$L_{eff} \approx (\alpha_P + \alpha_S / \cos \varphi)^{-1}, \tag{4}$$

$\alpha_P, \alpha_S$ are the linear extinction coefficients of the crystal and $\varphi$ is the angle between the signal photon direction and the back-normal to the slab, $\Delta \Omega_s = 2\pi \sin \theta \Delta \theta$ is the detector acceptance around $\vec{k}_S$, $\theta$ is approximately the deviation of one of PEPPE photon emission direction from the specular or from diffraction direction and $\theta_{Si}$ is the angle between the signal and idler photons or the opening angle between the signal and idler detectors.

The incidence angle onto the crystal of the pump photon must differ (see [5]) slightly by $\Delta\theta_{Br}$ from the Bragg angle $\theta_{Br}$:

$$\Delta\theta_{Br}=(\theta_S^2/2+3\chi_0)/\sin(2\theta_{Br})+\chi_0/\tan\theta_{Br}, \qquad (5)$$

Where $\chi_0$ is the linear susceptibility of the crystal. In [4,5] $\Delta\theta_{Br}$ is negative and of the order of (0.01 - 0.02) degrees.

The expression (1) differs from the corresponding formula (13) of [1] by the fact that in (13) it is given $dN_S/dt=AP_P\Delta\Omega$ where $P_P$ is the power of the pump photon beam. In [1] it is also numerically calculated the dependence of $dN_S/dt=AP_P\Delta\Omega$ upon $\hbar\omega_P$ (see Fig. 2 of [1]) in the good assumption $\omega_S\approx\omega_i\approx\omega_P/2$ for the (004) reflection of diamond, signal photon detector acceptance solid angle $\Delta\Omega_S=10^{-3}$ sr, at $P_P=10$ mW. In particular, when $\hbar\omega_P=$ 18.9 keV or $\lambda_P=0.65\ A°$ (this is the pump photon wavelength of the experiment [4]) $dN_S/dt=AP_P\Delta\Omega=0.78$ sec$^{-1}$ and, therefore, $A\approx 7.8\times10^4$ W$^{-1}$.s$^{-1}$sr$^{-1}$ or $A\approx 1.25\times10^{-14}$ eV$^{-1}$sr$^{-1}$. Note that at such wavelength $P_P=10$ mW is obtained if 3.3x10$^{12}$ pump photons per second hit the radiator, and this means that the probability $P_S^{1PP}$ of the detection of one entangled signal photon produced by one pump photon for the experimental parameters [1,4] is equal to $P_S^{1PP}$ = 2.4x10$^{-13}$. Taking into account the differences between the experimental parameters ($\theta_{Si}$, $\Delta\Omega_S$, detection efficiency, etc), this probability $P_S^{1PP}$ is in more or less satisfactory agreement with the results [4,5] of the coincidence experiments with the 3rd generation SR beams. Very close value for $P_S^{1PP}$ is obtained using the derived cross section per electron $d\sigma/d\Omega\approx(1/4\pi\alpha)r_e^4k_P^2\Delta x$, in which $r_e=e^2/mc^2$ is the classical electron radius, $\alpha=e^2/\hbar c$ is the fine structure constant and it is assumed that $\Delta x=\Delta\omega_P/\omega_P\approx 10^{-3}$ [5]. Let us note that using the Fig. 2 of [1] without complicated calculations of $G(hkl)$, $\vec{\theta}(spi)$ and other magnitudes, one can find $P_S^{1PP}(\omega_S)$ or $P_S^{1PP}(\omega_P)$, of course, in the same good assumption $\omega_S\approx\omega_i\approx\omega_P/2$ as function of $\lambda_P$ and obtain the curve with its maxima and zeros as the one in Fig. 2 of [1]. Let us also remind us that for measuring or calculating such dependence as given in Fig. 2 of [1], for each $\lambda_P$ the values of $\theta_{Br}$, $\Delta\theta_{Br}$, etc are different.

It is evident that such powers of pump photon beams can be provided also by the Williams-Weizsaker pseudophotons of high energy electrons, the spectral distribution of which for our approximation and for $\gamma=E/mc^2=1/\sqrt{1-\beta^2}\gg 1$ and $\hbar\omega_{WW}\ll\gamma mc^2$, according to [11], is equal to

$$\frac{dN_{WW}(\gamma\omega_{WW})}{d\omega_{WW}}\approx\frac{2\alpha}{\pi\omega_{WW}}\ln\left(\frac{\gamma mc^2}{\hbar\omega_{WW}}\right). \tag{6}$$

Using the Williams-Weizsaker method, the probability or the number $\Delta N_{PEPPE}(\omega_P)$ of PEPPE photons produced in a crystal by a single pseudophoton with frequency $\omega_P$ will be

$$\Delta N_{PEPPE}(\omega_P)=\frac{dN_{WW}(\gamma,\omega_P)}{d\omega_P}\Delta N_S(\omega_P). \tag{7}$$

Instead of integration of (5) over $\omega_{WW}$ in order to derive the number of PEPPE photons produced by the pseudophotons of a passing charged particle into the solid angle $\Delta\Omega_s=2\pi\sin\theta\Delta\theta$, we shall assume that in a narrow bandwidth $\Delta\omega_P$ the spectrum of pseudophotons is not varied. Therefore,

$$\Delta N_{PEPPE}^{1electron}(\gamma,\omega_P)=\int_{\omega_{WW}}\frac{dN_{WW}(\gamma,\omega_{WW})}{d\omega_{WW}}\Delta N_S(\omega_{WW})d\omega_{WW}\approx\\ \frac{dN_{WW}(\gamma,\omega_{WW})}{d\omega_{WW}}\Delta N_S(\omega_{WW})\Delta\omega_P \tag{8}$$

After all the substitutions one obtains

$$\Delta N_{PEPPE}^{1electron}(\gamma,\omega_P)=\frac{2\alpha A}{\pi}\ln\left(\frac{\gamma mc^2}{\hbar\omega_P}\right)(h\omega_P)\Delta\Omega\Delta x. \tag{9}$$

Just as the Darwin width in X-ray diffraction, $\Delta\omega_P$ and, therefore, $\Delta x$ must be small. However, this width is not determined only by the rocking width, but also by the size of the detector entrance (one can use also gradient crystal to increase this width). For further estimate we take $\Delta x\approx 10^{-3}$, though in [5] it is taken $\Delta x\approx 10^{-2}$. Like in [1,4], the spectral distributions of the photons are not considered; it is only assumed that $|\vec{k}_S|\approx|\vec{k}_i|$.

Before making estimates let us consider briefly the PEPPE arrangement shown in the following Fig.1.

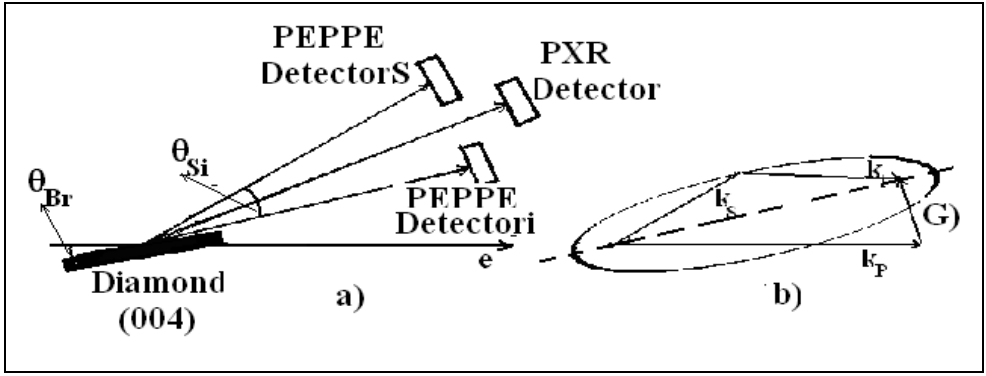

Fig. 1. The experimental arrangement (a) and the phase matching of the PEPPE (b).

In order to avoid the complicated calculations of the necessary parameters, especially the nonlinear X-ray susceptibility, we shall consider the same experimental parameters as in [1,4], i.e., we assume a 0.5 mm thick, symmetrically cut (004) diamond, photon energies $\hbar\omega_P \approx 19$ keV and $\hbar\omega_{S,i} \approx \hbar\omega_P/2 \approx 9.5$ keV, $\Delta\Omega_{S,i} \approx 10^{-3}$ sr, $\theta_{Br}$, $\theta_{Si}$ etc. Therefore, we can use the value of A from [1,4]. The 1 GeV electron beam with small cross section passes through a symmetrically cut diamond under incidence angle which is slightly smaller than the Bragg angle, $\theta_{Br}$. The crystal with thickness ~ 0.5 mm is slightly thicker than the necessary effective thickness [1,4] and has transversal sizes larger than the electron beam spot. The PXR detector placed at the specular direction for X-ray diffraction serves for monitoring purposes and detects the PXR photons produced with a probability ~$10^{-6}$-$10^{-5}$ per electron. Without taking into account the necessary small deviations from the Bragg angle the PEPPE detectors S and I, as in [1] and [4], are under angle ~1.5 degrees with respect to the specular direction. In coincidence they serve for detection of PEPPE signal and idler photons.

For the Mainz cw electron accelerator MAMI, the 850 MeV electron beam with average current 0.1 mA or electron number per second ~$6.2 \times 10^{14}$, taking $\Delta x = 10^{-3}$ with the help of (9) one expects $\Delta N_{PEPPE}^{1electron}(E_e = 850\,MeV, \hbar\omega_P = 19\,keV) \approx 1.1 \times 10^{-14}$ per electron, either 7.9 PEPPE photons per second or 28440 h$^{-1}$. This last yield is much better than it can be obtained on SR sources and is equal to the number expected for future X-ray FELs.

How could one perform the above estimates on the way of construction of more accurate theory of PEPPE? Following [12], one can write the probability or the cross section of the process multiplying more accurate expression for the pseudophoton spectrum by the X-ray reflection coefficients, say, for the Bragg geometry. The X-ray reflection coefficients without taking into account the non-linear part of the susceptibilities are well known (see, for instance, [13]). The building of the more correct PEPPE theory requires knowledge of these coefficients taking into account the real and imaginary parts of the non-linear susceptibilities and usage of more accurate solutions of Maxwell equations.

In conclusion, it is necessary to note: Firstly, as it is considered in [8], the X-ray microbunching of the electrons after the long SASE undulators [7], which can be measured before being spoiled by the method considered in [14], will make the proposed scheme of PEPPE a FEL for PEPPE due to the $\sim N_{mb}^2$ gain of the PEPPE yield, where $N_{mb}^2$ is the number of electrons in the microbunches. Secondly, in principle, PEPPE can be produced also when an electron passes through an amorphous plate, say, simultaneously with production of transition radiation (TR) photons [15]. Due to the proportionality of the TR photon yield to $Z^2$, the rate will be enhanced if instead of electrons one uses relativistic heavy ions with high Z values, i.e., beams at RHIC or LHC as in the proposal [16] for ring TRD. As in the above considered case of crystal radiator, in order to separate the production of PEPPE from the background in the case of TR one can measure the polarization of the two entangled photons in directions different from the TR polarization plane made of the direction of the primary particle and emission direction of the TR.